%% file: main.tex
\documentclass{article}

\usepackage{arxiv}

\usepackage{amssymb}
\usepackage{amsmath}
\usepackage{algorithmic}
\usepackage{textcomp}
\usepackage{xcolor}
\usepackage{subfig}
\usepackage[utf8]{inputenc} 
\usepackage[T1]{fontenc}    
\usepackage{hyperref}       
\usepackage{url}            
\usepackage{booktabs}       
\usepackage{amsfonts}       
\usepackage{nicefrac}       
\usepackage{microtype}      
\usepackage{cleveref}       
\usepackage{lipsum}         
\usepackage{graphicx}
\usepackage[numbers]{natbib}
\usepackage{doi}

\title{Propagation Constant Measurement Based on a Single Transmission Line Standard Using a Two-port VNA
}

\date{}

\author{
	Ziad Hatab, Arezoo Abdi, Gregor Steinbauer, Michael Ernst Gadringer, and Wolfgang Bösch \\
	Christian Doppler Laboratory for Technology Guided Electronic Component Design and Characterization\\
	Institute of Microwave and Photonic Engineering\\ 
	Graz University of Technology, Austria \\
	\texttt{\{z.hatab, arezoo.abdi, michael.gadringer, wbosch\}@tugraz.at} \thanks{Software implementation and measurements available online:\newline\url{https://github.com/ZiadHatab/two-port-single-line-propagation-constant}}
}

\renewcommand{\headeright}{}
\renewcommand{\undertitle}{}

\colorlet{mdtRed}{red!50!black}
\definecolor{myMagenta}{RGB}{236,0,140}
\definecolor{darkblue}{RGB}{0,49,110}
\hypersetup{pdfpagemode={UseOutlines},
	bookmarksopen=true,
	bookmarksopenlevel=0,
	hypertexnames=false,
	colorlinks=true,
	citecolor=myMagenta,
	linkcolor=black,
	urlcolor=mdtRed,
	pdfstartview={FitV},
	unicode,
	breaklinks=true,
}
\newcommand{\RE}[1]{\operatorname{Re}\left({#1}\right)}
\newcommand{\IM}[1]{\operatorname{Im}\left({#1}\right)}

\newcommand{\mat}[1]{\boldsymbol{#1}}
\newcommand{\bs}[1]{\boldsymbol{#1}}
\newcommand{\vc}[1]{\mathrm{vec}\left({#1}\right)}

\newcommand{\overbar}[1]{\mkern 1.5mu\overline{\mkern-1.5mu#1\mkern-1.5mu}\mkern 1.5mu}

\begin{document}
\maketitle

\begin{abstract}
	This study presents a new method for measuring the propagation constant of transmission lines using a single line standard and without prior calibration of a two-port vector network analyzer (VNA). The method provides accurate results by emulating multiple line standards of the multiline calibration method. Each line standard is realized by sweeping an unknown network along a transmission line. The network need not be symmetric or reciprocal, but must exhibit both transmission and reflection. We performed measurements using a slab coaxial airline and repeated the measurements on three different VNAs. The measured propagation constant of the slab coaxial airline from all VNAs is nearly identical. By avoiding disconnecting or moving the cables, the proposed method eliminates errors related to repeatability of connectors, resulting in improved broadband traceability to SI units.
\end{abstract}


\input{Sections/Section1}
\input{Sections/Section2}
\input{Sections/Section3}
\input{Sections/Section4}
\input{Sections/Section5}

\section*{Acknowledgment}
The financial support by the Austrian Federal Ministry for Digital and Economic Affairs and the National Foundation for Research, Technology, and Development is gratefully acknowledged. The authors also thank ebsCENTER for lending their Anritsu and Keysight VNAs, and Maury Microwave for their support in providing the airline cross-section dimensions of 8045P tuner.

\bibliographystyle{IEEEtran}
\bibliography{References/references.bib}

\end{document}

%% file: Sections/Section1.tex
\section{Introduction}
\label{sec:1}

The propagation constant is a critical parameter in transmission line analysis, providing valuable information about the electrical properties of materials at different frequencies. The need for accurate measurement of the propagation constant arises in various applications, such as material characterization \cite{Orend2022,Hatab2022a,Liu2019,Roelvink2013}, or estimation of the characteristic impedance of transmission lines, which allows impedance renormalization in various vector network analyzer (VNA) calibration methods \cite{Marks1991a}. Furthermore, knowledge of the propagation constant allows the analysis of losses along a transmission line, which is a critical aspect in signal integrity applications \cite{Bogatin2020,Zhang2010}. In general, there are many reasons to measure the propagation constant of guided wave structures such as transmission lines.

There are several methods to measure the propagation constant using a two-port VNA, but the most versatile method because of its broadband applicability is the multiline technique \cite{Marks1991}. In this method, multiple lines of different lengths are measured to sample the traveling wave along the line standards in a broadband scheme. However, this approach has several drawbacks, including the need for multiple lines, the possibility of uncertainties in their geometry, and the requirement for accurate repeated connection or probing, all of which contribute to measurement uncertainties \cite{DeGroot1996,Kaiser2000,Wong2013}.

To address some of the problems of the multiline method, some techniques have been introduced, such as the multireflect method \cite{Lewandowski2017,Popovic2020} and the line-network-network method \cite{Heuermann1997, Rolfes2003,Zhang2022,Orlob2013}. The multireflect method uses multiple identical reflect standards with different offsets to provide broadband measurement of the propagation constant. However, because it requires multiple identical independent standards, it is susceptible to repeatability errors due to repeated connection or probing. In addition, the propagation constant must be solved using optimization techniques that could diverge if not well conditioned. The line-network-network method involves moving an unknown symmetric and reciprocal network along a transmission line and solving for the propagation constant using the derived similarity equations. This method has limitations, such as the restriction to three offsets, which limits the frequency range, and the requirement to use symmetric and reciprocal offset networks. Results of relative effective permittivity measurements using this method were presented in \cite{Narayanan2014}, highlighting the sensitivity and limitations of this solution.

It is noteworthy that there is a significant amount of literature discussing the broadband measurement of the propagation constant using only two line standards of varying lengths, commonly known as the line-line method \cite{ReynosoHernandez2003,Koul2009,Fuh2013,Hasar2018}. Despite the different mathematical formulations used, all these methods are based on solving the characteristic polynomial of the eigenvalue problem associated with the thru-reflect-line (TRL) calibration \cite{Engen1979}. However, because of the use of only two lines, which often have a significant length difference to cover lower frequencies, the result of the propagation constant exhibits multiple resonance peaks, caused by integer multiples of half-wavelength occurrences in the electrical length of the transmission lines. To mitigate this issue, some authors have proposed post-processing techniques to filter the resonance peaks \cite{Fuh2012,RodriguezVelasquez2020}.

There are several indirect techniques for determining the propagation constant of transmission lines, which involve evaluating the permittivity of materials separately. These methods can be broadly classified into two categories: the resonant method and the transmission/reflection method. The resonant method, described in \cite{Stuchly1980,Kato2019}, estimates the permittivity from S-parameters at resonant frequencies, resulting in measurements only at specific frequencies. In contrast, the transmission/reflection method estimates the permittivity of a sample placed between two waveguides from the measured transmission and reflection coefficients. This method can be implemented using various configurations such as free space, rectangular waveguide, and coaxial line, as discussed in \cite{Kazemipour2015,BakerJarvis1990,Dudeck1992}.

In this paper, we present a different approach for measuring the propagation constant using a single transmission line standard, without the need for prior calibration of a two-port VNA. Our approach builds on the general idea introduced in \cite{Heuermann1997} by shifting an unknown network along a transmission line. Unlike the approach in \cite{Heuermann1997}, our method is not constrained by the number of offsets one can use, nor does the unknown network need to be symmetric or reciprocal. To combine all offset measurements, we propose a weighted $4\times4$ eigenvalue problem, inspired by the modified multiline method introduced in \cite{Hatab2022}. One of the key advantages of our proposed method is that it only requires a single transmission line to generate equations similar to those of the multiline method. Furthermore, highly repeatable measurements are possible because cable reconnection is not required. The remaining uncertainty is mainly due to the dimensional motion of the unknown network and the intrinsic noise of the VNA. We demonstrate the effectiveness of our method on a commercial slab coaxial airline tuner, where the offset network is the sliding tuning element. We performed measurements with three different VNA brands. The measured propagation constant obtained by the different VNAs show overlapping agreement. Our proposed approach offers a promising alternative to existing methods for measuring the propagation constant.

The remainder of this paper is structured as follows. In Section~\ref{sec:2}, we provide a detailed explanation of the mathematical derivation of the eigenvalue problem formulation that allows for the adaptation of the multiline method. Subsequently, in Section~\ref{sec:3}, we discuss the use of normalized eigenvectors to extract the complex exponential terms, which contain the propagation constant, and the utilization of least squares to derive an accurate estimate of the propagation constant. In Section~\ref{sec:4}, we describe the experimental setup, where we perform measurements using various VNAs and present the measured propagation constant of the slab coaxial airline, as well as a comparison with EM simulation. Finally, a conclusion is given in Section~\ref{sec:5}.


%% file: Sections/Section2.tex
\section{Formulating the eigenvalue problem}
\label{sec:2}

\begin{figure}[th!]
	\centering
	\includegraphics[width=.94\linewidth]{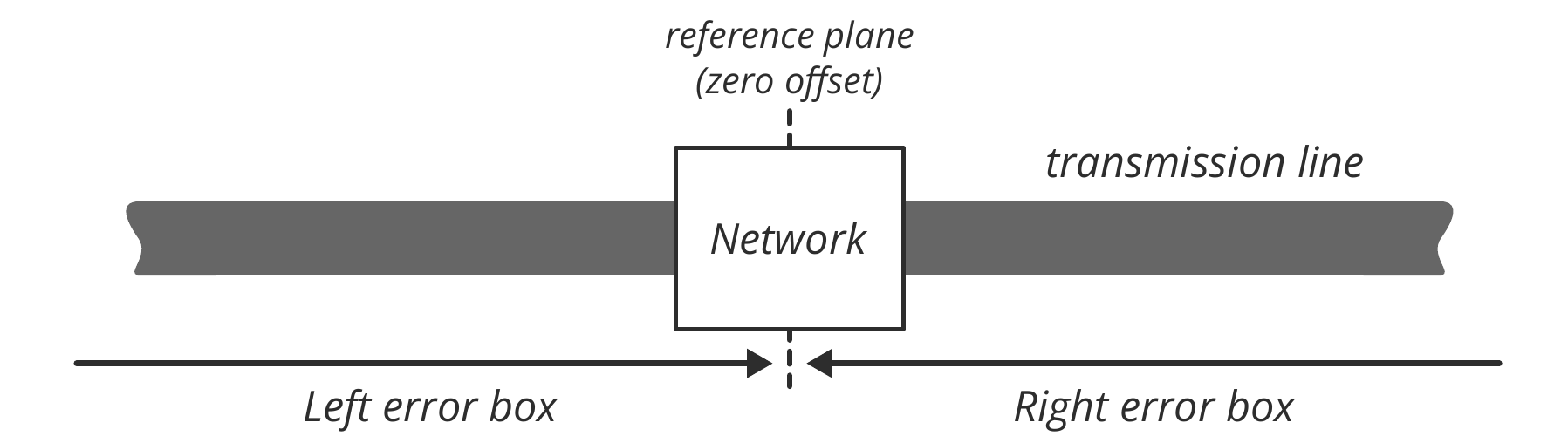}
	\caption{Illustration of network offset on a transmission line.}
	\label{fig:0}
\end{figure}

The general idea of the measurement setup is to move an unknown network along a transmission line. For each movement of the network, either to the left or to the right, we create two offset elements that are complementary to each other. When the offset length is zero, the offset elements are reduced to a thru connection, which we refer to as the reference plane. An illustration of this concept is shown in Figure~\ref{fig:0}. 

\begin{figure}[th!]
	\centering
	\includegraphics[width=.95\linewidth]{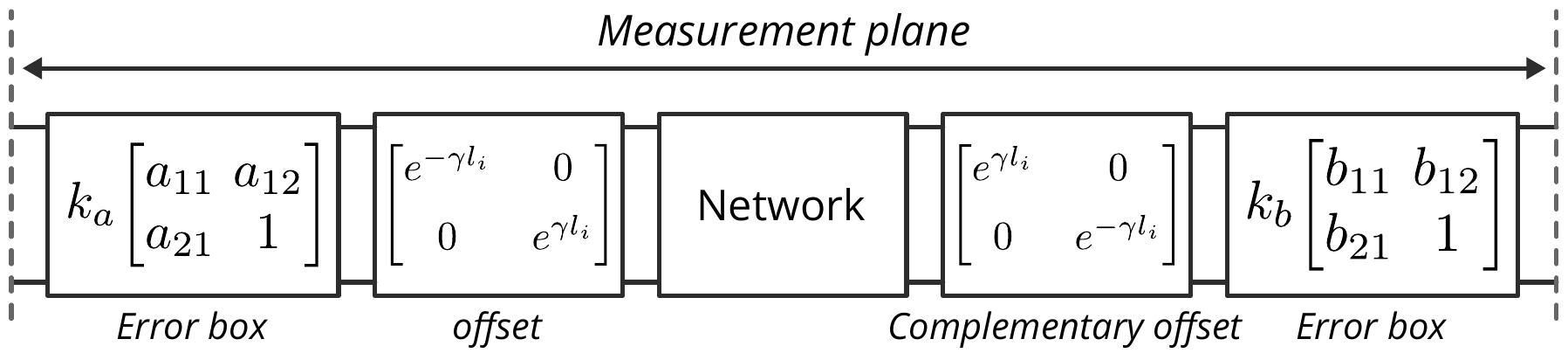}
	\caption{VNA two-port error box model of a network offsetted by a length $l_i$ (positive or negative). Each offset results in two complementary offset boxes. All blocks are given by their T-parameters.}
	\label{fig:1}
\end{figure}

Before proceeding with the mathematical derivation, we need to define the sign convention for the offset shift. In our analysis, we define that moving the network to the right results in a positive offset, while moving the network to the left results in a negative offset. This convention is shown in Figure~\ref{fig:1} as modeled by the error box model of a two-port VNA \cite{Marks1997}.

With the definition of the offsets in Figure~\ref{fig:1}, the measured T-parameters of the offsetted network by the length $l_i$ are given as follows:
\begin{equation}
	\bs{M}_i = \underbrace{k_ak_b}_{k}\underbrace{\left[\begin{matrix}a_{11} & a_{12}\\a_{21} & 1\end{matrix}\right]}_{\bs{A}}\bs{L}_i\bs{N}\bs{L}_i^{- 1} \underbrace{\left[\begin{matrix}b_{11} & b_{12}\\b_{21} & 1\end{matrix}\right]}_{\bs{B}},
	\label{eq:1}
\end{equation}
where $k$, $\bs{A}$, and $\bs{B}$ are the error terms of an uncalibrated two-port VNA. The matrices $\bs{L}_i$ and $\bs{N}$ are given as follows:
\begin{equation}
	\bs{L}_i = \begin{bmatrix} e^{-\gamma l_i} & 0 \\[5pt]
	 0 & e^{\gamma l_i} \end{bmatrix}, \quad \bs{N} = \begin{bmatrix} \frac{-S_{11}S_{22} + S_{12}S_{21}}{S_{21}} & \frac{S_{11}}{S_{21}} \\[5pt]
	 \frac{-S_{22}}{S_{21}} &\frac{1}{S_{21}} \end{bmatrix}.
	\label{eq:2}
\end{equation}

Here, $\gamma$ represents the propagation constant of the transmission line and $\{S_{11}, S_{12}, S_{21}, S_{22}\}$ are the S-parameters of the network $\bs{N}$. The S-parameters of the offset network are generally unknown, and the network can be asymmetric or non-reciprocal. However, the network must satisfy some basic criteria, which are listed below:
\begin{enumerate}
	\item All S-parameters must be non-zero within the considered frequency range ($|S_{ij}|>0$).
	\item The S-parameters of the network should not change as the network is moved.
	\item The network should not lead to the generation of additional modes along the transmission line.
\end{enumerate}

Although the first condition is unique to our method's formulation, the remaining two conditions are also similar to the multiline method \cite{Marks1991,Hatab2022}, which requires single-mode propagation and repeated error boxes. Fortunately, it is not difficult to design a system that satisfies these requirements. We will show this later in a Section~\ref{sec:4}, where we used a commercial sliding tuner that was not designed for our application, but met our conditions.

We now define the T-parameters of a new network by taking the difference in T-parameters of two offset networks of different lengths $l_i$ and $l_j$, which is given by
\begin{equation}
	\begin{aligned}	
	\overbar{\bs{N}}_{i,j} &= \bs{L}_i\bs{N}\bs{L}_i^{-1} - \bs{L}_j\bs{N}\bs{L}_j^{-1} \\
	&= \nu_{i,j}\begin{bmatrix}
		0 & \frac{S_{11}}{S_{21}}e^{-\gamma(l_i+l_j)} \\[5pt]
		\frac{S_{22}}{S_{21}}e^{\gamma(l_i+l_j)} & 0
	\end{bmatrix},
	\end{aligned}
	\label{eq:3}
\end{equation}
where 
\begin{equation}
	\nu_{i,j} = e^{-\gamma (l_i-l_j)} - e^{\gamma (l_i-l_j)}.
	\label{eq:4}
\end{equation}

The expression in \eqref{eq:3} is very similar to a line standard in multiline calibration, but now the line standard is described by an antidiagonal matrix and with additional multiplication factors. We define an equivalent measurement of a line standard by 
\begin{equation}
	\overbar{\bs{M}}_{i,j} = \bs{M}_i - \bs{M}_j = k\bs{A}\overbar{\bs{N}}_{i,j}\bs{B}.
	\label{eq:5}
\end{equation}

Similar to the multiline calibration, we also need an equation that describes the inverse of the measurements. This is given by
\begin{equation}
	\widehat{\overbar{\bs{M}}}_{i,j} = \bs{M}_i^{-1} - \bs{M}_j^{-1} = \frac{1}{k}\bs{B}^{-1}\widehat{\overbar{\bs{N}}}_{i,j}\bs{A}^{-1},
	\label{eq:6}
\end{equation}
where the matrix $\widehat{\overbar{\bs{N}}}_{i,j}$ is given by
\begin{equation}
	\begin{aligned}	
		\widehat{\overbar{\bs{N}}}_{i,j} &= \bs{L}_i^{-1}\bs{N}^{-1}\bs{L}_i - \bs{L}_j^{-1}\bs{N}^{-1}\bs{L}_j \\
		&= -\nu_{i,j}\begin{bmatrix}
			0 & \frac{S_{11}}{S_{12}}e^{-\gamma(l_i+l_j)} \\[5pt]
			\frac{S_{22}}{S_{12}}e^{\gamma(l_i+l_j)} & 0
		\end{bmatrix}.
	\end{aligned}
	\label{eq:7}
\end{equation}

Given the expressions in \eqref{eq:5} and \eqref{eq:6}, we can construct an eigenvalue problem in terms of $\bs{A}$ as follows:
\begin{equation}
	\overbar{\bs{M}}_{i,j}\widehat{\overbar{\bs{M}}}_{n,m} = \bs{A}\overbar{\bs{N}}_{i,j}\widehat{\overbar{\bs{N}}}_{n,m}\bs{A}^{-1},
	\label{eq:8}
\end{equation}
where the matrix product $\overbar{\bs{N}}_{i,j}\widehat{\overbar{\bs{N}}}_{n,m}$ is given by 
\begin{equation}
	\overbar{\bs{N}}_{i,j}\widehat{\overbar{\bs{N}}}_{n,m} = -\kappa\nu_{i,j}\nu_{n,m}\begin{bmatrix}
		e^{-\gamma(l_{i,j}^+-l_{n,m}^+)} & 0 \\
		0 & e^{\gamma(l_{i,j}^+-l_{n,m}^+)}
	\end{bmatrix},
	\label{eq:9}
\end{equation}
with 
\begin{equation}
	\kappa = \frac{S_{11}S_{22}}{S_{21}S_{12}}, \qquad l_{i,j}^+ = l_i + l_j, \qquad l_{n,m}^+ = l_n + l_m.
	\label{eq:10}
\end{equation}

To have a valid eigenvalue problem, we need at least three unique offsets, where one of the offsets $l_n$ or $l_m$ can be equal to $l_i$ or $l_j$, but $l_i \neq l_j$, or vice versa. However, with three offsets, we have three possible pairs of eigenvalue problems. In fact, for $N \geq 3$ offsets, we have $N(N-2)(N^2-1)/8$ possible pairs of eigenvalue problems. This is because for a set of $N$ offsets, we have $N(N-1)/2$ pairs, and when we create pairs from $N(N-1)/2$ pairs, we substitute the equation into itself, resulting in $N(N-2)(N^2-1)/8$ pairs of pairs. 

To address the issue of multiple eigenvalue problems, we refer to our previous work in \cite{Hatab2022,Hatab2023}, where a similar problem was presented in the context of multiline calibration. This problem was solved by combining all measurements using a weighting matrix, reducing the problem to solving a single $4 \times 4$ eigenvalue problem, regardless of the number of lines. This method not only reduced the size of the problem, but also allowed us to express both error boxes $\bs{A}$ and $\bs{B}$ simultaneously in a single matrix using Kronecker product notation. By applying the techniques described in \cite{Hatab2022,Hatab2023}, we obtain the following set of equations:
\begin{subequations}
	\begin{align}
		\overbar{\bs{M}} &= k\bs{X}\overbar{\bs{N}},\label{eq:11a}\\[5pt] 
		\widehat{\overbar{\bs{M}}}^{T}\bs{P} &= \frac{1}{k}\widehat{\overbar{\bs{N}}}^{T}\bs{P}\bs{X}^{-1},\label{eq:11b}
	\end{align}
	\label{eq:11}
\end{subequations}
with, 
\begin{subequations}
	\begin{align}
		\bs{X} &= \bs{B}^T\otimes\bs{A},\label{eq:12a}\\[5pt]
		\overbar{\bs{M}} &= \begin{bmatrix} \vc{\overbar{\bs{M}}_{1,2}} & \cdots & \vc{\overbar{\bs{M}}_{i,j}} \end{bmatrix},\label{eq:12b}\\[5pt]
		\overbar{\bs{N}} &= \begin{bmatrix} \vc{\overbar{\bs{N}}_{1,2}} & \cdots & \vc{\overbar{\bs{N}}_{i,j}} \end{bmatrix},\label{eq:12c}\\[5pt]
		\widehat{\overbar{\bs{M}}} &= \begin{bmatrix} \vc{\widehat{\overbar{\bs{M}}}_{1,2}} & \cdots & \vc{\widehat{\overbar{\bs{M}}}_{i,j}} \end{bmatrix},\label{eq:12d}\\[5pt]
		\widehat{\overbar{\bs{N}}} &= \begin{bmatrix} \vc{\widehat{\overbar{\bs{N}}}_{1,2}} & \cdots & \vc{\widehat{\overbar{\bs{N}}}_{i,j}} \end{bmatrix},\label{eq:12e}\\[5pt]
		\bs{P} &= \begin{bmatrix}
			1 & 0 & 0 & 0\\
			0 & 0 & 1 & 0\\
			0 & 1 & 0 & 0\\
			0 & 0 & 0 & 1
		\end{bmatrix},\quad\text{where, } \bs{P}=\bs{P}^{-1}=\bs{P}^T.\label{eq:12f}
	\end{align}
	\label{eq:12}
\end{subequations}

Details on the definition and properties of the Kronecker product ($\otimes$) and the matrix vectorization ($\vc{}$) can be found in the reference \cite{Brewer1978}. 

We now formulate the main eigenvalue problem by defining a new matrix $\bs{W}$, which we multiply on the right side of \eqref{eq:11a}. We call this matrix the weighting matrix. In the next step, we construct the weighted eigenvalue problem by multiplying the new equation on the left side of \eqref{eq:11b}. This results in
\begin{equation}
	\underbrace{\overbar{\bs{M}}\bs{W}\widehat{\overbar{\bs{M}}}^{T}\bs{P}}_{\bs{F}} = \bs{X}\underbrace{\overbar{\bs{N}}\bs{W}\widehat{\overbar{\bs{N}}}^{T}\bs{P}}_{\bs{H}}\bs{X}^{-1}.
	\label{eq:13}
\end{equation}

The expression presented in \eqref{eq:13} represents a similarity problem between the matrices $\bs{F}$ and $\bs{H}$, with $\bs{X}$ as the transformation matrix. The purpose of introducing the weighting matrix $\bs{W}$ is to transform this similarity problem into an eigenvalue problem by forcing $\bs{H}$ into a diagonal form. It turns out that if $\bs{W}$ is any non-zero skew-symmetric matrix, then $\bs{H}$ takes a diagonal form \cite{Hatab2022}. However, we do not only want to diagonalize $\bs{H}$, but also want to maximize the distance between the eigenvalues, which in turn minimizes the sensitivity in the eigenvectors \cite{Wilkinson1988}. For multiline calibration, the optimal form of $\bs{W}$ was derived in \cite{Hatab2022}, and since the formulation in \eqref{eq:13} is similar to that discussed in \cite{Hatab2022}, we use the same choice of $\bs{W}$ with some scaling modifications. The optimal weighting matrix $\bs{W}$ can be written as follows, taking into account the scaling factors:
\begin{equation}
	\bs{W}^H = -\kappa(\bs{z}\bs{y}^T - \bs{y}\bs{z}^T),
	\label{eq:14}
\end{equation}
where
\begin{subequations}
	\begin{align}
		\bs{y}^T &= \begin{bmatrix}
			\nu_{1,2}e^{\gamma l_{1,2}^+} & \ldots & \nu_{i,j}e^{\gamma l_{i,j}^+}
		\end{bmatrix},\\
		\bs{z}^T &= \begin{bmatrix}
			\nu_{1,2}e^{-\gamma l_{1,2}^+} & \ldots & \nu_{i,j}e^{-\gamma l_{i,j}^+}
		\end{bmatrix}.
	\end{align}
	\label{eq:15}
\end{subequations}

As a result of choosing $\bs{W}$ as defined in \eqref{eq:14}, the expression in \eqref{eq:13} takes an eigendecomposition form, as given below.
\begin{equation}
	\bs{F} = \bs{X} \begin{bmatrix}
		0 & 0 & 0 & 0\\
		0 & \lambda & 0 & 0\\
		0 & 0 & -\lambda & 0\\
		0 & 0 & 0 & 0
	\end{bmatrix}\bs{X}^{-1},
	\label{eq:16}
\end{equation}
where $\lambda$ is real-valued and proportional to the square Frobenius norm of the matrix $\bs{W}$, given by
\begin{equation}
	\lambda = \frac{1}{2}\left\|\bs{W}\right\|_F^2 = \frac{1}{2}\sum_{i,j} |w_{i,j}|^2.
	\label{eq:17}
\end{equation}

There are two ways to compute $\bs{W}$, the first is the direct method where we already know the propagation constant $\gamma$ and the factor $\kappa$ which describes the unknown network. Naturally, the first option is not practical since both $\gamma$ and $\kappa$ are unknown. The better option is to apply a rank-2 Takagi decomposition to the left side of the following equation, as described in \cite{Hatab2023} for multiline calibration.
\begin{equation}
	\underbrace{\widehat{\overbar{\bs{M}}}^{T}\bs{P}\overbar{\bs{M}}}_\text{measurement} = \underbrace{\widehat{\overbar{\bs{N}}}^{T}\bs{P}\overbar{\bs{N}}}_\text{model}.
	\label{eq:a1}
\end{equation}

Note that the left side of \eqref{eq:a1} contains only the measurement data, while the right side describes the model. Also, the error boxes are not present in \eqref{eq:a1}. To determine $\bs{W}$, we need to calculate the rank-2 Takagi decomposition. This is done in two steps. First, we compute the rank-2 of \eqref{eq:a1} via signular value decomposition (SVD), and then we apply the Takagi decomposition to decompose the matrix into its symmetric basis \cite{Chebotarev2014}. This looks as follows:
\begin{equation}
	\widehat{\overline{\bs{N}}}^{T}\bs{P}\overline{\bs{N}} = \underbrace{s_1\mat{u}_1\mat{v}_1^H + s_2\mat{u}_2\mat{v}_2^H}_\text{rank-2 SVD from measurement} = \underbrace{\mat{G}\mat{G}^T}_\text{Takagi}
	\label{eq:a2}
\end{equation}

Then, the weighting matrix is given by 
\begin{equation}
	\bs{W}^H = \pm \bs{G}\begin{bmatrix}
		0 & j\\ -j & 0
	\end{bmatrix}\bs{G}^T
	\label{eq:a3}
\end{equation}

The derivation process of the matrix $\bs{W}$ is described in more detail in \cite{Hatab2023}. To resolve the sign ambiguity, one approach is to select the answer that has the smallest Euclidean distance to a known estimate. Such as an estimate can be obtained from an approximate knowledge of the material properties of the transmission line.

The last step is the solution of the eigenvectors described by $\bs{X}$ in \eqref{eq:16}. The solution of the eigenvectors has been discussed in \cite{Hatab2022}. It is worth noting that we cannot solve the matrix $\bs{X}$ uniquely, but only up to a diagonal matrix multiplication. Therefore, to define a unique solution for $\bs{X}$, we normalize its columns so that the diagonal elements are equal to one. This is written as follows:
\begin{equation}
	\widetilde{\bs{X}} = \bs{X}\mathrm{diag}(a_{11}b_{11},b_{11},a_{11},1)^{-1},
	\label{eq:18}
\end{equation}
where $a_{11}$ and $b_{11}$ are part of the error boxes $\bs{A}$ and $\bs{B}$ (see \eqref{eq:1} and \eqref{eq:12a}).
 

%% file: Sections/Section3.tex
\section{Least squares solution for the propagation constant}
\label{sec:3}

Knowing $\widetilde{\bs{X}}$ from the eigenvector solutions, we can extract the complex exponential terms that contain the propagation constant. To do this, we first multiply the inverse of the normalized error terms to all vectorized measurements of the offset network. This is given by
\begin{equation}
	\bs{E} = \widetilde{\bs{X}}^{-1}\bs{M} = \mathrm{diag}(ka_{11}b_{11},kb_{11},ka_{11},1)\bs{N}^\prime,
	\label{eq:19}
\end{equation}
where
\begin{subequations}
	\begin{align}
		\bs{M} &= \begin{bmatrix} \vc{\bs{M}_1} & \cdots & \vc{\bs{M}_N} \end{bmatrix},\label{eq:20a}\\[5pt]
		\bs{N}^\prime &= \begin{bmatrix} \vc{\bs{L}_1\bs{N}\bs{L}_1^{- 1}} & \cdots & \vc{\bs{L}_N\bs{N}\bs{L}_N^{- 1}}\end{bmatrix}.\label{eq:20b}
	\end{align}
	\label{eq:20}
\end{subequations}

Since we do not know the remaining error terms $k,a_{11},b_{11}$, as well as the S-parameters of the network $\bs{N}$, we need to choose a reference offset to eliminate these unknown factors. For simplicity, we choose the first offset, which we define as zero, i.e., $l_1=0$ (any other choice is also valid). As a result, the positive and negative complex exponential terms are given as follows, using indexing notation based on Python.
\begin{subequations}
	\begin{align}
		\bs{E}[1,1:]/\bs{E}[1,0] =& \begin{bmatrix}
			e^{2\gamma l_2} & e^{2\gamma l_3} & \ldots & e^{2\gamma l_N}
		\end{bmatrix},\label{eq:21a}\\[5pt]
		\bs{E}[2,1:]/\bs{E}[2,0] =& \begin{bmatrix}
			e^{-2\gamma l_2} & e^{-2\gamma l_3} & \ldots & e^{-2\gamma l_N}
		\end{bmatrix}.\label{eq:21b}
	\end{align}
	\label{eq:21}
\end{subequations}

Now that we have the complex exponential terms, we can extract the exponents using the complex logarithm function and determine $\gamma$ using the least squares method, while taking care of any phase unwrapping. First, since we have both the positive and negative complex exponential terms, we can account for both by averaging them. This is done by defining a new vector $\bs{\tau}$:
\begin{equation}
	\bs{\tau} = \begin{bmatrix} \frac{e^{2\gamma l_2} + 1/e^{-2\gamma l_2}}{2} & \cdots & \frac{e^{2\gamma l_N} + 1/e^{-2\gamma l_N}}{2} \end{bmatrix}^T.
	\label{eq:22}
\end{equation}

The next step is to calculate the logarithm to extract the exponents, which is given by
\begin{equation}
	\bs{\phi} = \log\left( \bs{\tau}\right) + j2\pi \bs{n}, \quad \text{where, } \bs{n}\in\mathbb{Z}^{N-1}.
	\label{eq:23}
\end{equation}

The phase unwrapping factor $\bs{n}$ can be estimated by rounding the difference between $\bs{\phi}$ and an estimated value. This is given by
\begin{equation}
	\bs{n} = \mathrm{round}\left(\frac{\IM{\bs{\phi}} - 2\gamma_\mathrm{est}\bs{l}}{2\pi}\right),
	\label{eq:24}
\end{equation}
where $\gamma_\mathrm{est}$ is a known approximation for $\gamma$, and $\bs{l}$ is a vector containing all length offsets except the reference zero offset. The initial estimate for $\gamma_\mathrm{est}$ can be derived from the material properties of the transmission line. 

Finally, we can determine $\gamma$ through weighted least squares \cite{DeGroot2002}, 
\begin{equation}
	\gamma = \frac{\bs{l}^T\bs{V}^{-1}\bs{\phi}}{\bs{l}^T\bs{V}^{-1}\bs{l}},
	\label{eq:25}
\end{equation}
where $\bs{V}^{-1}$ is given by
\begin{equation}
	\bs{V}^{-1} = \bs{I}_{(N-1)\times(N-1)} - \frac{1}{N}\bs{1}_{N-1}\bs{1}^T_{N-1},
	\label{eq:26}
\end{equation}

The matrix $\bs{I}$ is the identity matrix and $\bs{1}$ is a vector of ones. The weighting matrix $\bs{V}^{-1}$ is necessary because each measurement has a common reference, which is $l_1$. Therefore, the correlation between the measurements had to be taken into account by the matrix $\bs{V}^{-1}$ \cite{DeGroot2002}.

\begin{figure}[th!]
	\centering
	\includegraphics[width=0.8\linewidth]{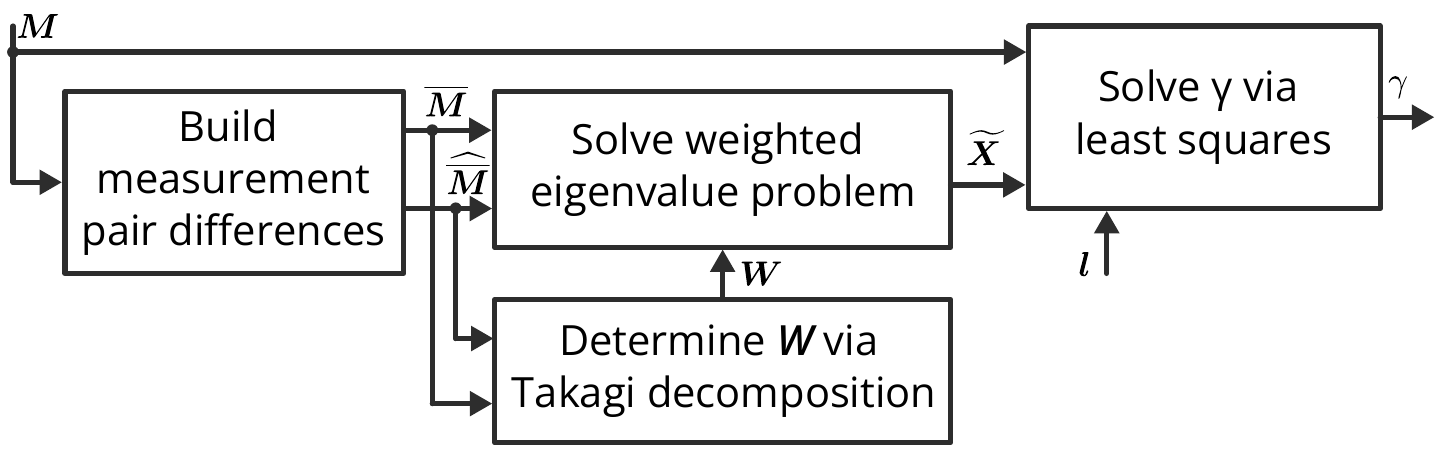}
	\caption{Block diagram summary of the proposed propagation constant measurement method. The matrix $\bs{M}$ contains the T-parameter measurements of all offsets. The vector $\bs{l}$ contains the relative length of the offsets with respect to the reference offset (i.e., the zero offset).}
	\label{fig:2}
\end{figure}

Figure~\ref{fig:2} summarizes the mathematical derivation presented in this and the previous sections, and provides a visual representation of the steps taken to compute the propagation constant.


%% file: Sections/Section4.tex
\section{Experiment and Discussion}
\label{sec:4}

\subsection{Measurement setup}
\begin{figure}[th!]
	\centering
	\includegraphics[width=0.8\linewidth]{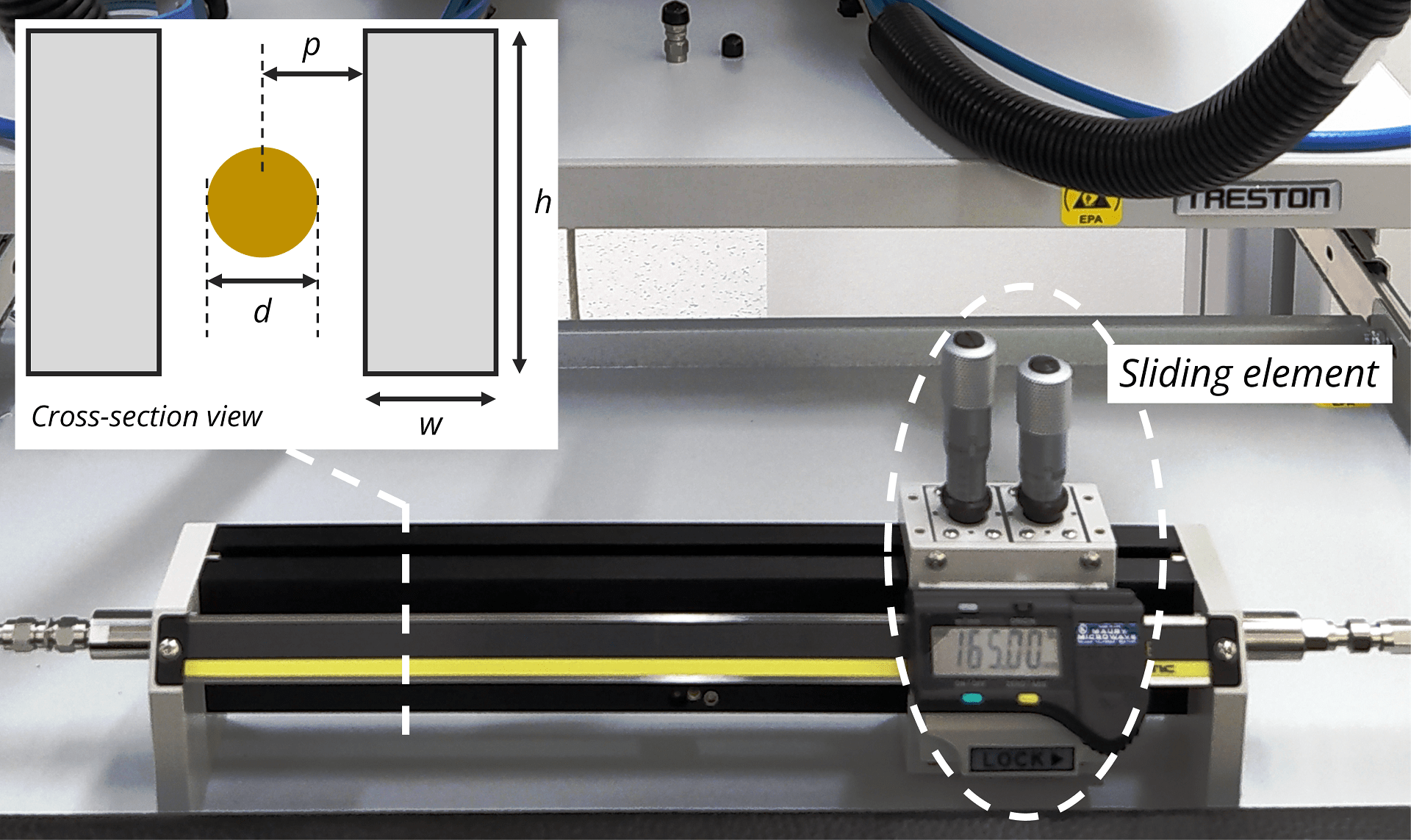}
	\caption{Maury Microwave 8045P tuner. The cross-section dimensions of the airline are given as follows: $w=9.398\,\mathrm{mm}$, $h=40.691\,\mathrm{mm}$, $d=3.040\,\mathrm{mm}$, and $p=2.778\,\mathrm{mm}$.}
	\label{fig:3}
\end{figure}

For demonstration purposes, we used the slide screw tuner 8045P from Maury Microwave as an implementation of the offset network, where the transmission line is a slab coaxial airline that supports frequencies up to 18\,GHz. The tuner is depicted in Figure~\ref{fig:3}.

\begin{figure}[th!]
	\centering
	\includegraphics[width=.95\linewidth]{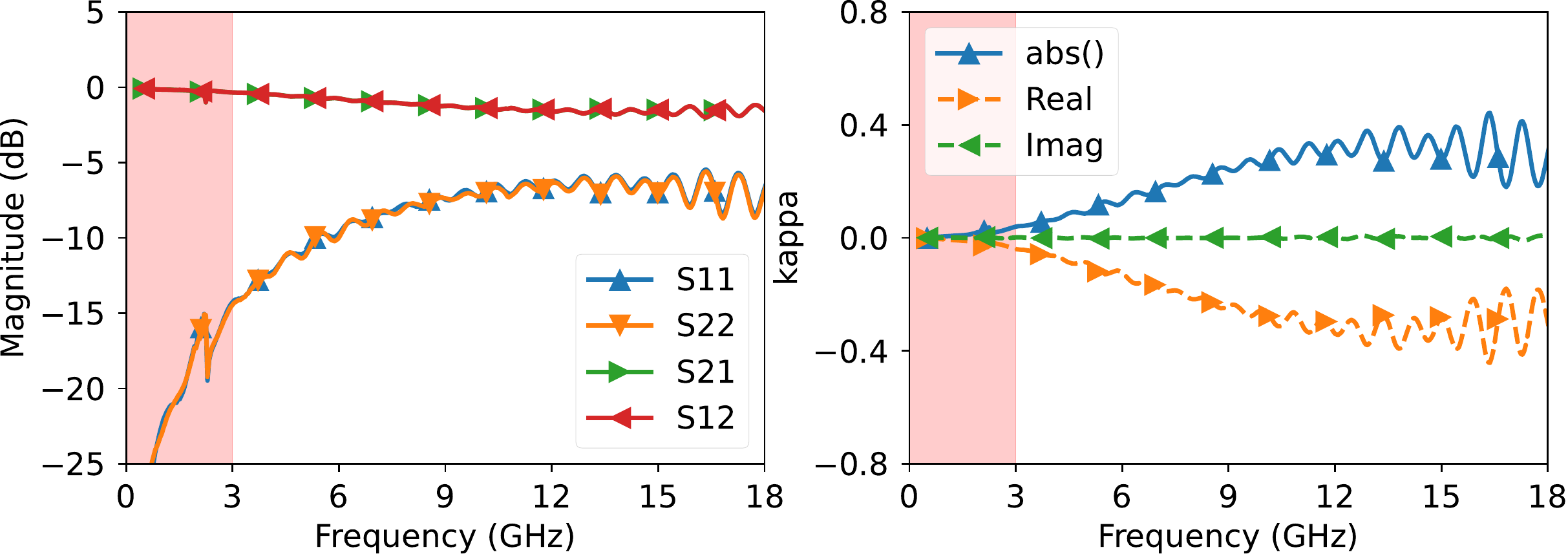}
	\caption{Calibrated measurement of the tuner after tuning. The highlighted frequency range below 3\,GHz is not usable due to small reflection and resonance.}
	\label{fig:4}
\end{figure}

For our method to work, we require that the unknown network (i.e., the tuner element) be both reflective and transmissive, as the factor $\kappa$ in \eqref{eq:9} can explode to infinity if the network is only reflective, and can be zero if the network is only transmissive. Ideally, we want $\kappa=1$ to minimize its effect on the eigenvalue problem. However, we also want to avoid scenarios where the network causes the generation of additional modes or resonances. Therefore, we adjusted the tuner with an already calibrated VNA to tune the network to a desired response, as shown in Figure~\ref{fig:4}. It should be noted that this step of tuning the tuner with an existing calibrated VNA is only necessary because the tuner is a commercial product designed for circuit matching applications and not for our purpose. If we were designing the network ourselves, we would not need to measure it with a calibrated VNA because we would have already designed it to meet our frequency specifications. Also, the S-parameters of the network are never explicitly used in the derivation of the propagation constant.

As shown in Figure~\ref{fig:4}, we set the lower frequency to 3\,GHz to avoid very low return loss and resonances. We then measured the airline using different uncalibrated VNAs. This was done to demonstrate that even if we changed the measurement system, we would still get consistent results because the error boxes would not be affected by uncertainties caused by connector and cable movement. For the offset lengths, we chose $\{0,21,66,81,84,93,117,123,171,192\}\,\mathrm{mm}$, which ensures that the eigenvalue $\lambda$ in \eqref{eq:16} does not reach zero in the target frequency range.

\begin{figure}[th!]
	\centering
	\subfloat[]{\includegraphics[width=.32\linewidth]{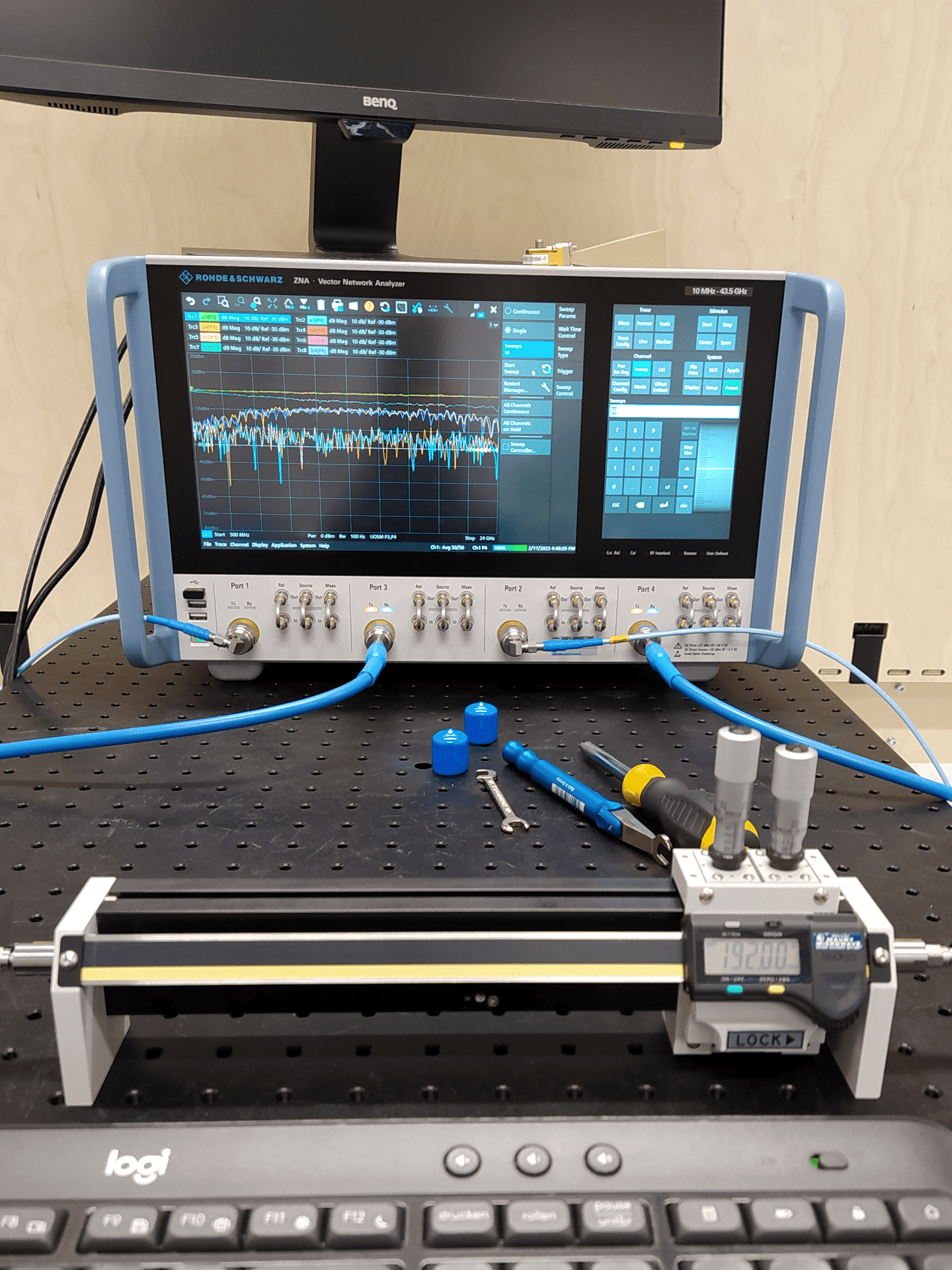}
			\label{fig:4xq}}~
	\subfloat[]{\includegraphics[width=.32\linewidth]{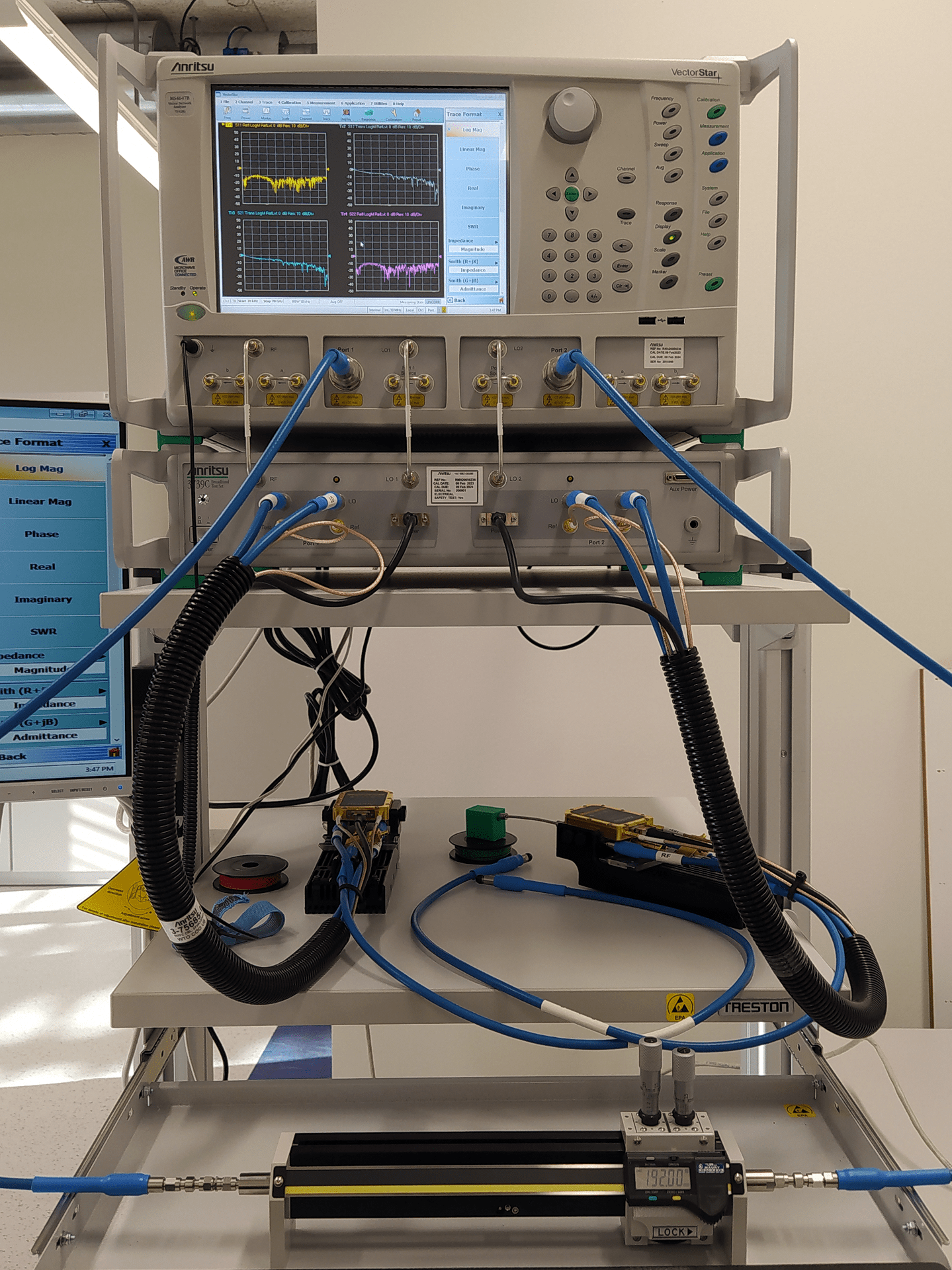}
			\label{fig:4xb}}~
	\subfloat[]{\includegraphics[width=.32\linewidth]{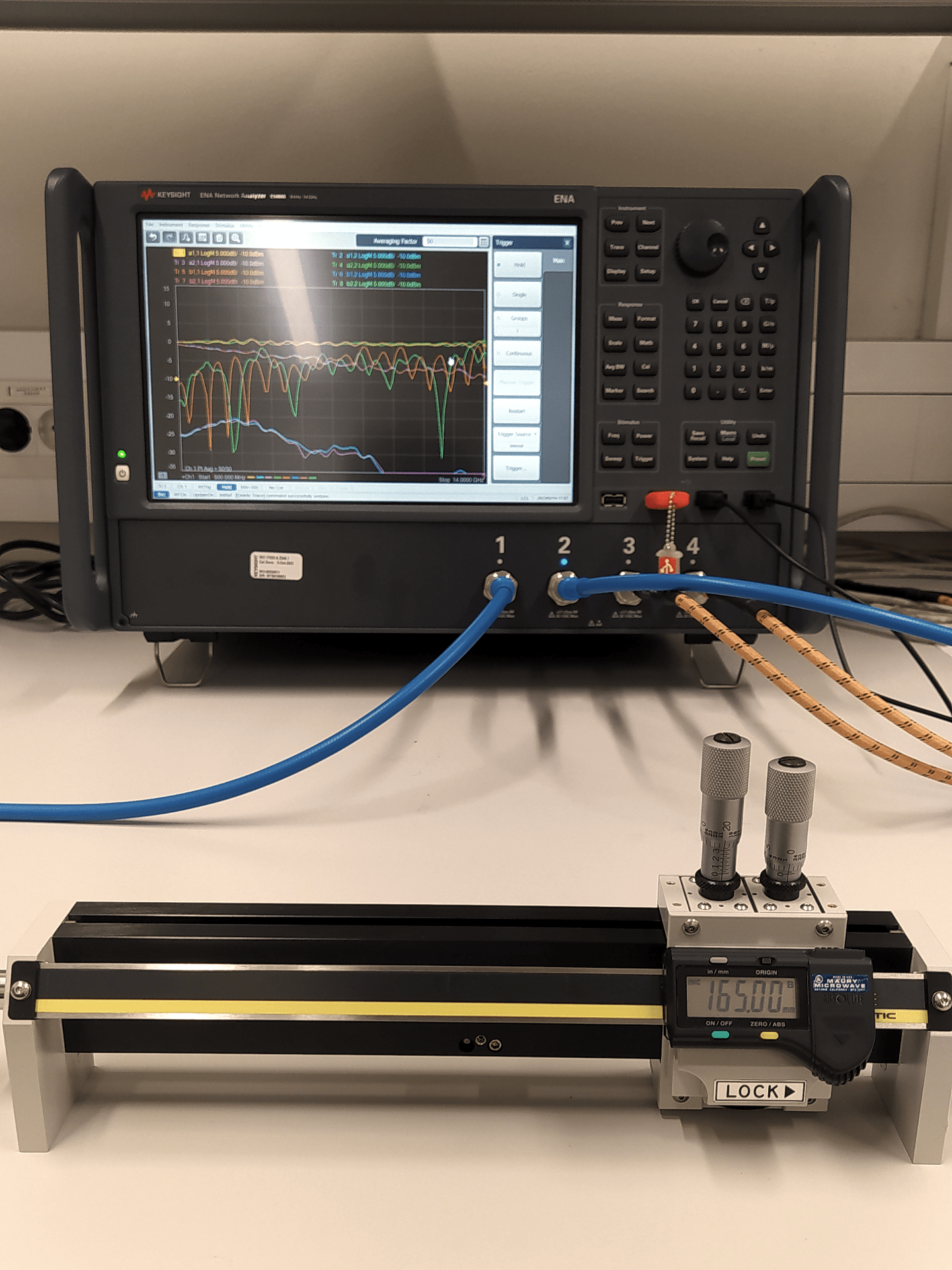}
			\label{fig:4xc}}
	\caption{The VNAs used for the measurements. (a) Rohde \& Schwarz ZNA, (b) Anritsu VectorStar, and (c) Keysight ENA.}
	\label{fig:4x}
\end{figure}

The VNAs used for the measurements are: Anritsu VectorStar, R\&S ZNA and Keysight ENA. The ENA is limited to 14\,GHz. All VNAs were placed in the same room to provide the same room conditions. The power level and IF bandwidth for all VNAs were set to 0\,dBm and 100\,Hz, respectively. Due to the low loss of the airline, an average measurement of 50 frequency sweeps was calculated to reduce noise. Pictures of the three instruments are shown in Figure \ref{fig:4x}.

\subsection{Results and discussion}

All measurements of the different offsets were taken without prior calibration of the VNAs. The collected data is then read in Python using the \textit{scikit-rf} package \cite{Arsenovic2022}. In Figure~\ref{fig:4x} we show the measured magnitude response of $S_{11}$ and $S_{21}$ from all three VNAs for the offset $123\,\mathrm{mm}$. From the figure, we can see that all three VNAs give different responses because the error boxes are different for each VNA.
\begin{figure}[th!]
	\centering
	\includegraphics[width=0.95\linewidth]{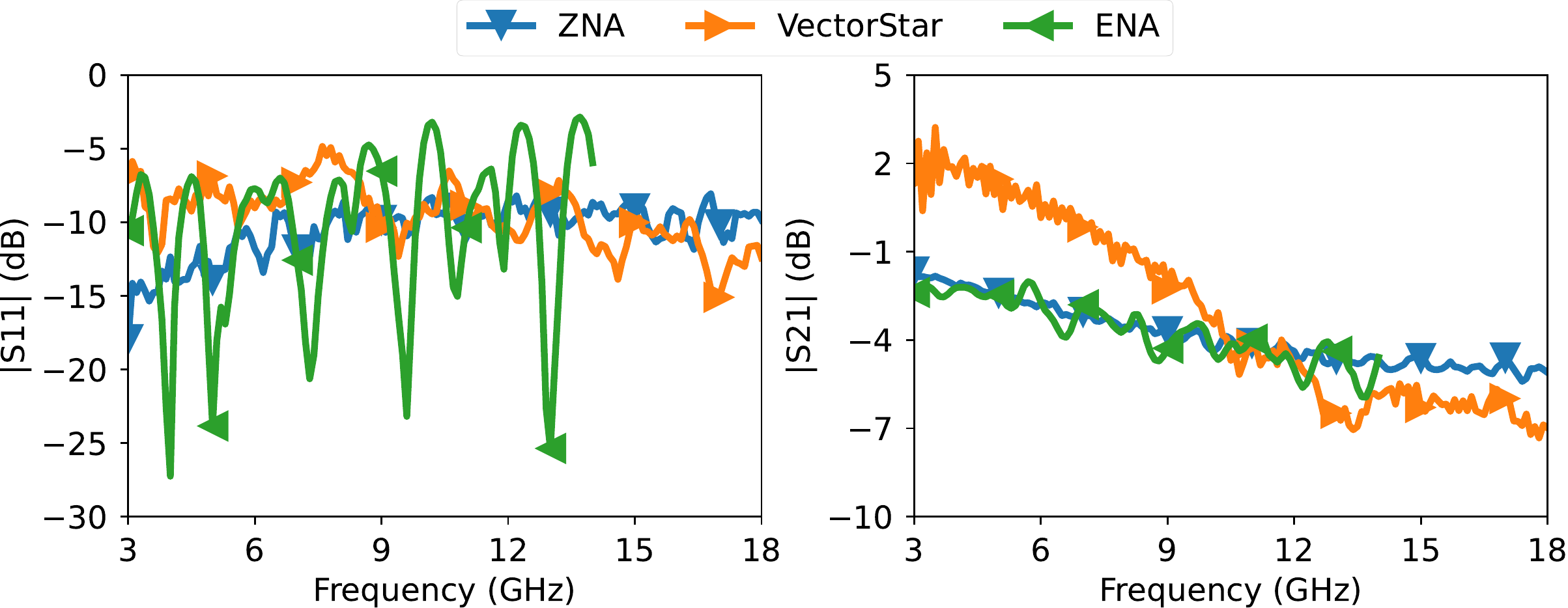}
	\caption{Raw measurements from the three VNAs of the magnitude response of $S_{11}$ and $S_{21}$ of the 8045P tuner, at an offset location of $123\,\mathrm{mm}$.}
	\label{fig:4xx}
\end{figure}

\begin{figure}[th!]
	\centering
	\includegraphics[width=0.95\linewidth]{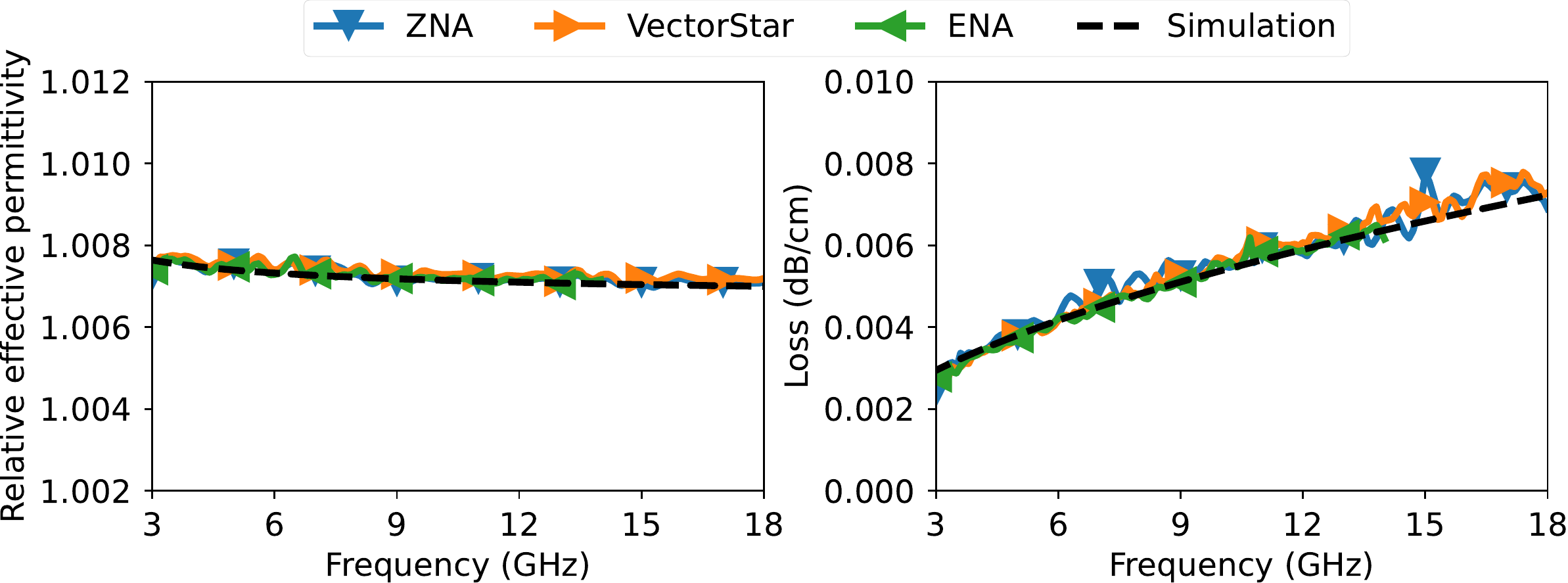}
	\caption{Extracted measurement of relative effective permittivity and loss per unit length of the slab coaxial airline, as well as EM simulated results for an anodic coating of $15\,\mu\mathrm{m}$ and inner conductor conductivity of $35\%$ IACS.}
	\label{fig:5}
\end{figure}

After collecting all raw measurements for all the offsets and from all the VNAs, the data were processed to extract the propagation constant according to the discussion in Sections \ref{sec:3} and \ref{sec:4}. For easier and better interpretation of the extracted propagation constant, we have plotted in Figure~\ref{fig:5} the real part of the relative effective permittivity and the loss per unit length of the slab coaxial airline from all three VNA measurements. The real part of the relative effective permittivity and the loss per unit length are calculated from the propagation constant as follows:
\begin{equation}
	\epsilon_{\mathrm{r,eff}}^{\prime} = -\RE{\left( \frac{c_0\gamma}{2\pi f}\right)^2}\quad (\text{Unitless}), \qquad \text{loss} = \frac{20\times10^{-2}}{\ln10}\RE{\gamma}\quad (\text{dB/cm}),
	\label{eq:27}
\end{equation}
where $c_0$ is the speed of light in vacuum and $f$ is the frequency.

The relative effective permittivity and loss per unit length results presented in Figure~\ref{fig:5} show clear agreement between all VNA measurements, demonstrating the high repeatability of the proposed method even when using different VNA setups. 

We also performed an EM simulation with the dimensional parameters of the airline given in Figure~\ref{fig:3}. Unfortunately, we did not have information on the metal types of the inner and outer conductors. From the appearance of the inner conductor, we think it is made of some kind of brass. For the ground plates, we think they are made of aluminum because they have a black anodized coating, which is typical for aluminum components. The anodized layer is often based on aluminum oxide and typically has a relative permittivity of $8.3$ \cite{Argall1968}. Since the thickness of the oxide layer and the exact conductivity of brass are unknown, we ran some values for the thickness of the anodic layer and the conductivity of brass. We found that a coating thickness of $15\,\mu\mathrm{m}$ and a relative conductivity of $35\%$ IACS (International Annealed Copper Standard) overlap with the measurement shown in Figure~\ref{fig:5}. The value obtained for the thickness of the anodic layer is quite typical to obtain a dark black coating \cite{HENLEY1982}. The conductivity of the inner conductor of $35\%$ IACS (=20.3\,MS/m) is within the range of common brass types \cite{Davis2001}.

The purpose of the simulation is to show that the results obtained from the proposed method of measuring the propagation constant do indeed translate into realistic properties of the transmission line. In fact, with the proposed method, one could characterize materials in reverse, as in our case, the conductivity of the metal.

Another aspect that may be of interest is the quality of the extracted propagation constant by varying the length and number of offsets. In the results shown in Figure~\ref{fig:5} we used 10 offsets ranging from 0 to 192\,mm. Now we consider different cases. These cases are listed in Table~\ref{tab:1}.
\begin{table}[th!]
	\centering
	\caption{Considered cases of different offset lengths.}
	\label{tab:1}
	\begin{tabular}{@{$\quad$}c|c@{$\quad$}}
		\toprule
		Cases       & Offset lengths (mm)              \\ \midrule
		Case 1      & $0,21,81$                          \\
		Case 2      & $0,21,192$                         \\
		Case 3      & $0,21,66,117,192$                  \\
		Case 4      & $0,21,81,93,117,123,192$           \\
		All offsets & $0,21,66,81,84,93,117,123,171,192$ \\ \bottomrule
	\end{tabular}
\end{table}

\begin{figure}[th!]
	\centering
	\includegraphics[width=0.95\linewidth]{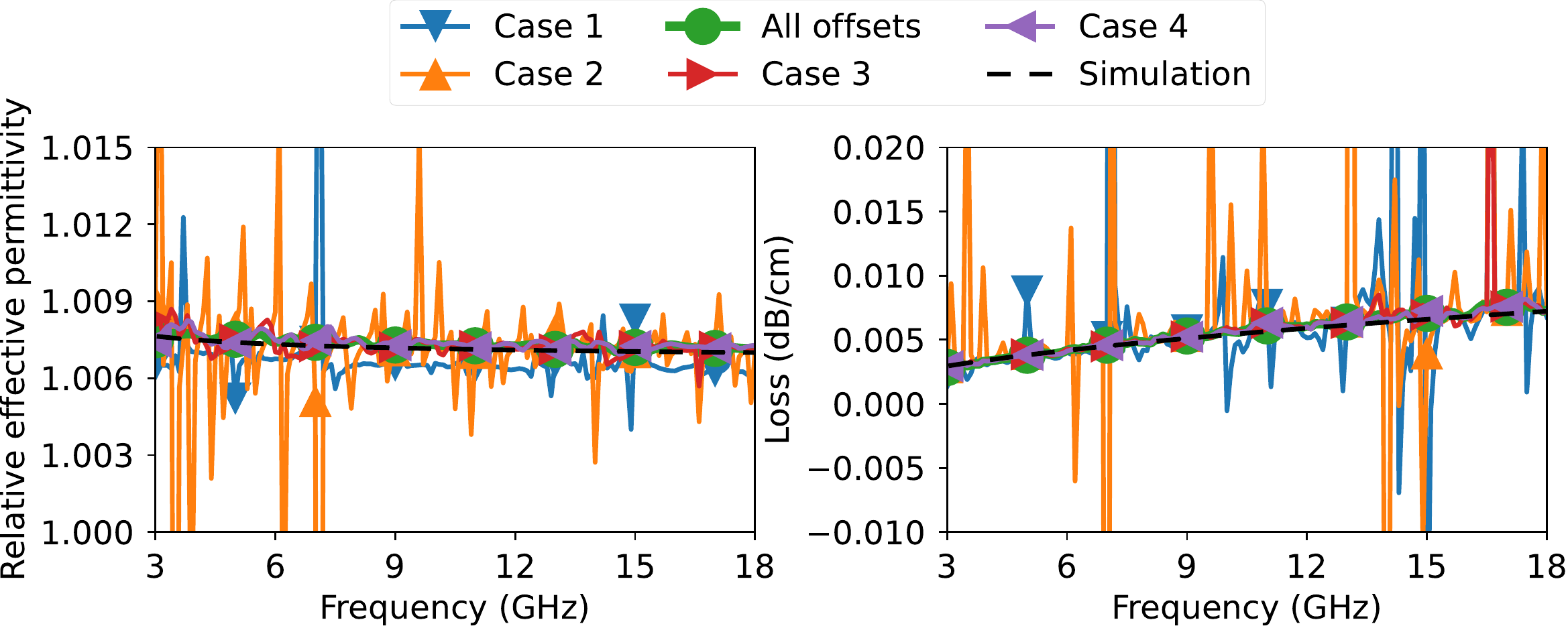}
	\caption{Extracted measurement of relative effective permittivity and loss per unit length of the slab coaxial airline for various combinations of offsets from the VectorStar VNA measurements.}
	\label{fig:5x}
\end{figure}

\begin{figure}[th!]
	\centering
	\includegraphics[width=0.85\linewidth]{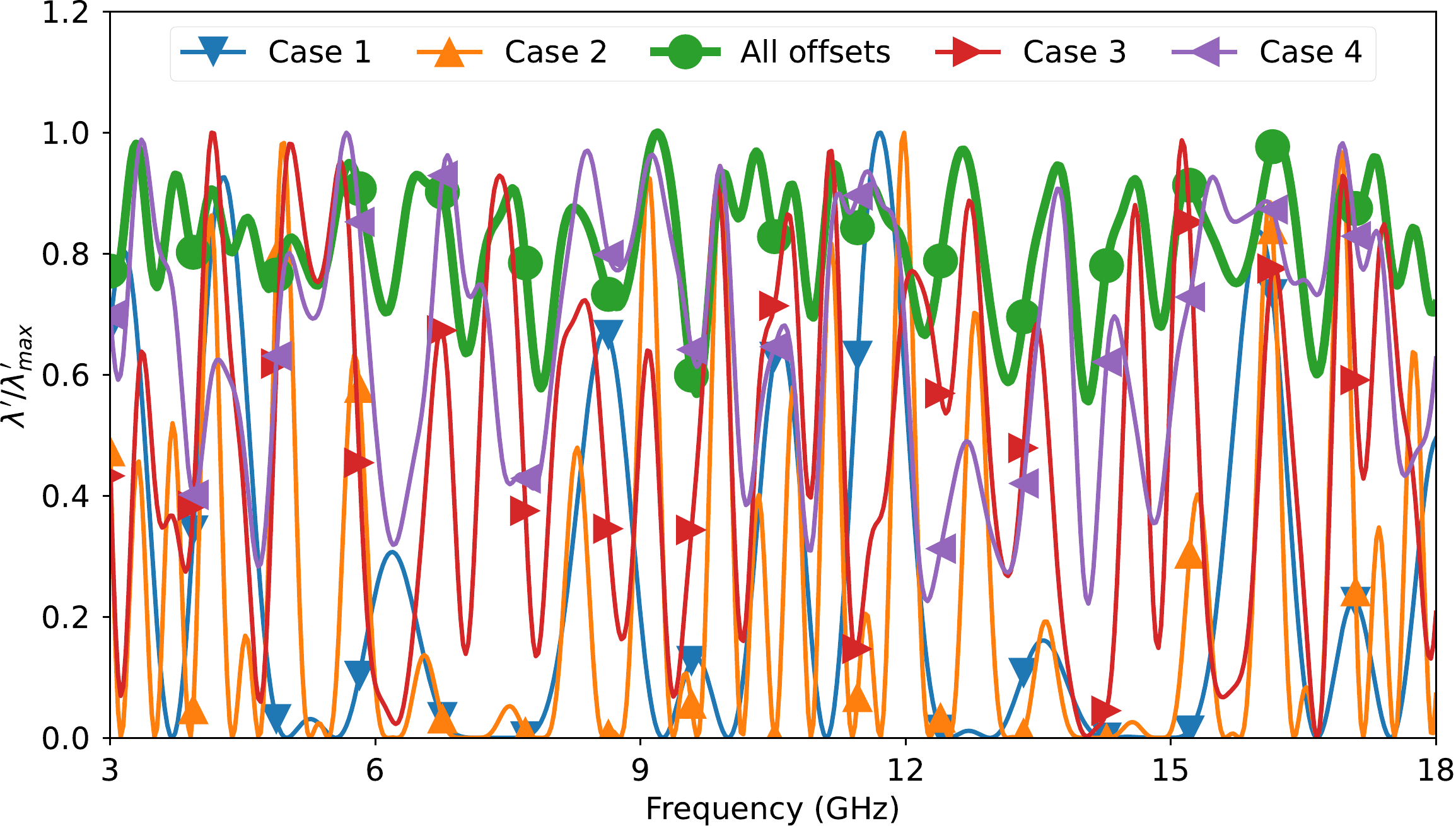}
	\caption{The scaled eigenvalue normalized to its maximum for the invested offset length cases.}
	\label{fig:5xx}
\end{figure}

In Figure~\ref{fig:5x} we show the results of the relative effective permittivity and the loss per unit length of the slab coaxial airline from the VectorStar VNA measurements for all the cases mentioned in Table~\ref{tab:1}. Cases 1 and 2 show the results when only three offsets are considered. Case 2 differs from Case 1 in that we have replaced the last offset with a much longer offset. The results of both cases 1 and 2 are poor and show multiple resonances. For case 2 we see more resonances than for case 1. This is the result of the eigenvalue crossing zero at multiple frequencies (see Figure~\ref{fig:5xx}). In case 3 we spread the offsets further to include 5 offsets. We can see a clear improvement over cases 1 and 2. We can further improve the accuracy of the extracted relative effective permittivity and loss per unit length by further spreading the offsets as in case 4, where we use 7 offsets. In case 4, we obtain results of similar accuracy to the case of using all 10 offset lengths.

The quality of the extracted propagation constant depends on the eigenvalue $\lambda$ as defined in \eqref{eq:16}. As the eigenvalue approaches zero, the eigenvectors become more sensitive, which in turn affects the calculation of the extracted propagation constant. To visualize the differences between different scenarios, we present a scaled representation of the eigenvalue $\lambda$ for each case. This scaled representation excludes the influence of the network through the common factor $\kappa$, which is invariant over all offset lengths. Since $\kappa>0$ was established earlier, variations in the eigenvalues can only be induced by the choice of offset lengths. Accordingly, we define the normalized version of the eigenvalue by dividing it by the absolute value of $\kappa$, as shown below:
\begin{equation}
	\lambda = \frac{1}{2}\left\|\bs{W}\right\|_F^2 = \frac{|\kappa|^2}{2}\left\|\bs{z}\bs{y}^T - \bs{y}\bs{z}^T\right\|_F^2 \quad\Longrightarrow\quad \lambda^\prime = \frac{\lambda}{|\kappa|^2} = \frac{1}{2}\left\|\bs{z}\bs{y}^T - \bs{y}\bs{z}^T\right\|_F^2
	\label{eq:28}
\end{equation}

In Figure~\ref{fig:5xx}, we present a plot of the scaled eigenvalue normalized to its maximum value, which facilitates a consistent comparison as the number of offsets varies. As illustrated in the figure, for cases 1 and 2, the eigenvalue exhibits multiple zero crossings at various frequencies. Similarly, in case 3, the eigenvalue approaches zero at several instances, although to a lesser extent than in cases 1 and 2. In contrast, in case 4, the eigenvalue never reaches zero, but attains values closer to zero at specific frequencies than when all 10 offsets are utilized. Ideally, a flat eigenvalue over frequency would be preferred, but this would necessitate employing even more offsets. This is not different from the multiline calibration approach proposed in \cite{Hatab2022}, where a finer spacing between lines results in a flatter eigenvalue over frequency. Therefore, utilizing a broader range of offset lengths is highly advantageous for enhancing the accuracy of results across frequency. It is also noteworthy that the eigenvalue possesses a bandpass characteristic, whereby the lowest and highest frequency limits are bound by the largest and smallest relative offset, respectively. 

For comparison, it is worth noting that the multiline method necessitates the measurement of multiple line standards of different lengths, a process that can introduce errors due to connector repeatability. Achieving high repeatability in this context poses a significant mechanical challenge, especially concerning connectors, and automating this process represents an even greater hurdle. In contrast, our proposed method eliminates the need for physical contact between the sliding element and the transmission line. Furthermore, although the sliding process is performed manually in the example presented, it could be automated by employing a linear actuator, thus eliminating the need for any user interaction with the measurement system.


%% file: Sections/Section5.tex
\section{Conclusion}
\label{sec:5}

We presented a new broadband method for measuring the propagation constant of transmission lines that does not require the prior calibration of a two-port VNA or the use of multiple line standards. This method provides accurate results by emulating the use of multiple line standards through sweeping an unknown network along a transmission line. The shifted network does not have to be symmetric or reciprocal, but it must exhibit both transmission and reflection properties and remain invariant when moved along the line. The experimental results obtained using different VNAs on a slab coaxial airline with a slider tuner showed consistent agreement with each other and with EM simulation.

One of the significant advantages of the proposed method is that it uses the same eigenvalue formulation as multiline calibration, but without the need for disconnecting or moving the cables. As a result, it eliminates errors related to connector repeatability and provides improved broadband traceability to the SI units. Moreover, since the offsets are implemented by simply moving the unknown network laterally, the process can be easily automated using an automated linear actuator. Therefore, the proposed method can accurately measure the propagation constant without requiring any physical interaction from the user on the measurement system.